\documentclass[prb,twocolumn,superscriptaddress,showpacs,amsmath,amssymb]{revtex4-1}

\usepackage{graphicx}
\usepackage{latexsym}
\usepackage{amsmath}
\usepackage{amssymb}
\usepackage{amsfonts}
\usepackage{color}
\usepackage{bm}
\usepackage{verbatim}

\definecolor{MS-color}{RGB}{128,0,128}

\newcommand{\ak}[1]{{\color{magenta} #1}}

\usepackage{framed}
\definecolor{shadecolor}{RGB}{222,222,221}

\begin{document}

\title{Thermally induced spin torque and domain wall motion in superconductor/antiferromagnetic insulator bilayers}

\author{G. A. Bobkov}
\affiliation{Moscow Institute of Physics and Technology, Dolgoprudny, 141700 Russia}

\author{I. V. Bobkova}
\affiliation{Institute of Solid State Physics, Chernogolovka, Moscow reg., 142432 Russia}
\affiliation{Moscow Institute of Physics and Technology, Dolgoprudny, 141700 Russia}
\affiliation{Dubna State University, Dubna,  141980, Russia}

\author{A. M. Bobkov}
\affiliation{Institute of Solid State Physics, Chernogolovka, Moscow reg., 142432 Russia}

\author{Akashdeep Kamra}
\affiliation{Center for Quantum Spintronics, Department of Physics, Norwegian University of Science and Technology, NO-7491 Trondheim, Norway}

\date{\today}


\begin{abstract}
We theoretically investigate domain wall motion in an antiferromagnetic insulator layer caused by thermally generated spin currents in an adjacent spin-split superconductor layer. An uncompensated antiferromagnet interface enables the two crucial ingredients underlying the mechanism - spin splitting in the superconductor and absorption of spin currents by the antiferromagnet. Treating the superconductor using the quasiclassical theory and the antiferromagnet via Landau-Lifshitz-Gilbert description, we find domain wall propagation along the thermal gradient with relatively large velocities $\sim 100$ m/s. Our proposal exploits the giant thermal response of spin-split superconductors in achieving large spin torques towards driving domain wall and other spin textures in antiferromagnets.
\end{abstract}

\maketitle

\section{Introduction}

Recent advancements in stabilizing and manipulating textured magnetic states has led to a paradigmatic transition in the role of magnets in futuristic solid state devices~\cite{Gobel2020,Yu2020}. In addition to being passive memory storage elements, spin textures may allow for an active participation of magnets in data processing~\cite{Han2019}. These continued advancements, however, rely on realizing effective methods to manipulate these spin textures, such as a domain wall (DW). A wide range of methods, from external magnetic field~\cite{Schryer1974} to thermal spin transfer torque~\cite{Ralph2008,Jen1986,Jiang2013,Bauer2012,Hatami2007,Hinzke2011,Slonczewski2010}, have been considered as candidates controlling DW motion in ferromagnets. The finite net magnetization in a ferromagnet turns out crucial in nearly all of these mechanisms. On the other hand, antiferromagnet (AFs) offer various advantages due to their fundamentally different and faster dynamics~\cite{Gomonay2014,Baltz2018,Jungwirth2018,Jungwirth2016,Gomonay2010}, but lack net magnetization and the associated easy control. In the context of DW motion, AFs support significantly larger DW velocities thereby offering a faster operation of devices~\cite{Yu2020,Yuan2020,Kim2017}.

Temperature gradient as a drive for DW motion has gained fresh impetus on account of several novel spin-thermal effects discovered in the last years~\cite{Bauer2012,Hatami2007,UchidaSSE,Meyer2017,Gomonay2018}. Thermal gradient encompasses a broad range of mechanisms that could induce DW motion such as electronic spin current generation~\cite{Hatami2007}, magnonic spin currents~\cite{Yan2011,Slonczewski2010,Hinzke2011}, entropic spin torques~\cite{Wang2014,Schlickeiser2014,Selzer2016,Donges2020} and so on. Furthermore, there are competing processes at play within these mechanisms resulting in a complex interplay. For example, magnons in a ferromagnet may push a DW away on reflection via linear momentum delivery~\cite{Yan2013}. Alternately, they may pull the DW in the direction of their origin on transmission and angular momentum delivery~\cite{Yan2011}. This competition between pull and push forces is still more complex for AFs due to a varying spin of the magnons~\cite{Tveten2014,Yu2018}. On the other hand, entropic torques tend to drive DWs towards the hotter end in both ferro- and antiferromagnets~\cite{Schlickeiser2014,Selzer2016,Donges2020}. Hence, thermally induced spin torques and DW motion in (antiferro)magnets constitutes a subject with intriguing physics~\cite{Gomonay2018,Yang2019}, in addition to a high technological relevance. 

Various thermoelectric (spin) effects in normal or magnetic metals are small because they scale as temperature divided by the Fermi energy, with latter being a large quantity~\cite{Bauer2012,Machon2013,Ozaeta2014}. This smallness of thermoelectric effects can be overcome in superconductors where the superconducting gap, or equivalently the critical temperature, replaces the Fermi energy as the relevant parameter~\cite{Machon2013,Ozaeta2014}. Thus, giant thermoelectric effects and thermal spin currents can be achieved in hybrids comprising superconductor (S) and ferromagnet (F) layers~\cite{Machon2013,Machon2014,Ozaeta2014,Kolenda2016,Kolenda2017,Kolenda2016_2}. The two key ingredients in achieving such giant effects are spin-splitting of quasiparticle density of states in the superconductor~\cite{Tedrow1986,Hao1991} and spin-resolved transport. Both of these are accomplished in S/F hybrids~\cite{Bergeret2018,Heikkila2019}. As a result, a giant thermally induced quasiparticle spin current and DW velocities in the latter are possible and have recently been predicted~\cite{Bobkova2020}. The recent prediction of spin-splitting induced in S by an adjacent AF insulator bearing an uncompensated interface~\cite{Kamra2018} raises the question if a similar thermal spin current and DW motion can be realized in S/AF hybrids, which constitutes the subject of this paper. Such uncompensated moments at AF surfaces have been observed in numerous experiments~\cite{Zhang2011,Kappenberger2003,Sampaio2003,Camarero2006,Roy2005,Valev2006,Ohldag2003,Blomqvist2004,Mathieu1998}. Furthermore, a variety of effects that are highly sensitive to the nature of the interface with an AF have been predicted recently~\cite{Hellman2017,Rabinovich2019,Erlandsen2019,Johansen2019,Johnsen2020}.

We theoretically investigate an S/AF bilayer in which the AF hosts a DW and bears an uncompensated interface with S such that only one of the two sublattices in AF is exposed to S~\cite{Kamra2017,Kamra2018}. This results in a finite and spatially varying exchange field in S. We find that subjecting the hybrid to a thermal gradient primarily results in a large quasiparticle spin current in S. The latter exerts spin torque on the AF DW and moves it along the direction of the thermal gradient with velocities $\sim 100$ m/s. We evaluate the thermally generated spin currents in S microscopically using quasiclassical theory and treat the dynamics in AF using two-sublattice Landau-Lifshitz-Gilbert description. Besides numerically analyzing the ensuing response and DW velocities in a broad parameter space, we also derive analytic expressions in the limit of small DW width. Our analysis further provides guidance to experiments in optimizing the spin-splitting and thermal gradients that realize the highest DW velocities.

\section{Model}
\begin{figure}[tb]
\begin{center}
   \includegraphics[width=82mm]{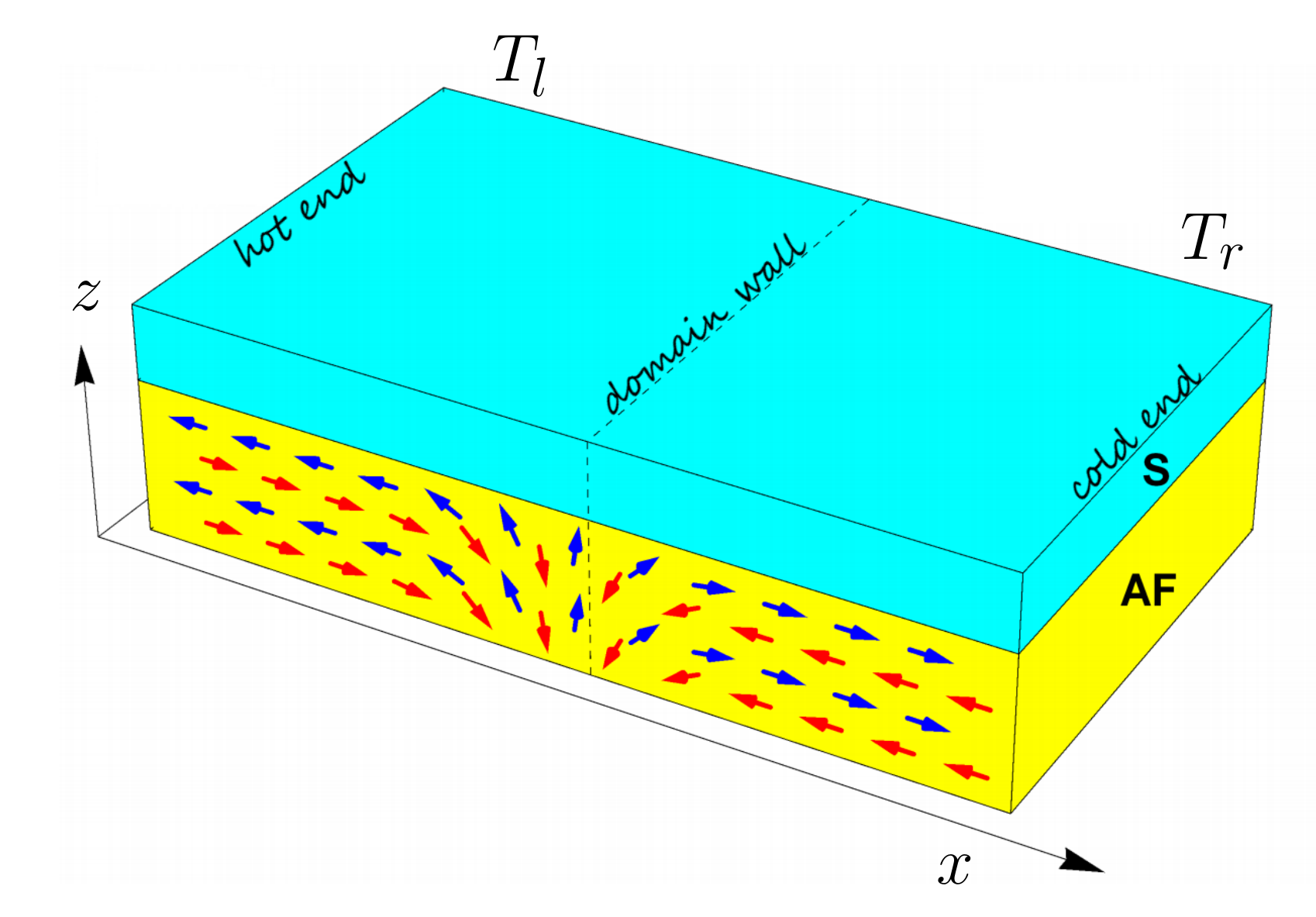}
      \caption{System under consideration: superconductor/antiferromagnet (S/AF) bilayer with uncompensated magnetic moment at the S/AF interface. Blue arrows depict spin of the A sublattice atoms, while red arrows correspond to the B sublattice. The temperature difference $T_l-T_r$ is applied along the $x$-direction.}
 \label{sketch}
 \end{center}
 \end{figure}

The model system that we consider is shown in Fig.~\ref{sketch}. It is a thin film bilayer consisting of a spin-textured antiferromagnet interfaced to a spin-singlet superconductor. The antiferromagnet is assumed to be an insulator with an uncompensated magnetic moment at the S/AF interface, that is the interface possesses finite magnetization. It has also been predicted that the uncompensated surface magnetization should appear at interfaces of magnetoelectric AFs~\cite{Belashchenko2010} and in the presence of Rashba spin-orbit coupling~\cite{Belashchenko2010,Lund2020}.

It has been demonstrated~\cite{Kamra2018} that if the thickness of the S film $d_S$ is smaller than the superconducting coherence length $\xi_S$, the magnetic proximity effect, that is the influence of the adjacent antiferromagnet on the S film can be described by adding the effective exchange field to the quasiclassical Eilenberger equation, which we use below to treat the superconductor. While in general the proximity effect at S/AF interfaces is not  reduced to the effective exchange only \cite{Kamra2018}, in the framework of the present study we neglect other terms, which can be viewed as additional spin-flip scattering due to magnetic impurities, and focus on the effect of the spin texture. 

The bilayer film is assumed to be connected to reservoirs having different temperatures $T_{l,r}$. Under these conditions, the presence of the effective exchange field in the superconductor results in appearance of a spin torque acting on the antiferromagnet. The physical nature of the torque is related to the giant quasiparticle spin Seebeck effect in the spin-split superconductor~\cite{Machon2013,Machon2014,Ozaeta2014}, which pumps quasiparticle spin into the superconducting region in the  vicinity of the DW. The pumped spin interacts with the AF interface magnetization via the interface exchange coupling. The mechanism is in complete analogy to spin torque arising at the superconductor/ferromagnet interface~\cite{Bobkova2020}. However, here we investigate how this torque influences the dynamics in {\it antiferromagnets}, which, in general, strongly differs from the dynamics in ferromagnets.

\subsection{Magnetization dynamics in the antiferromagnet}
The Landau-Lifshitz-Gilbert (LLG) equation can be written for each sublattice separately~\cite{Kamra2018_2}:
\begin{eqnarray}
\frac{\partial\bm m_i}{\partial t} = -\gamma \bm m_i \times \bm H_{eff}^i + \sum \limits_j \alpha_{ij} \bm m_i \times \frac{\partial\bm m_j}{\partial t} + \bm N_i,~~~~~~
\label{LLG}
\end{eqnarray}
where $\bm m_i = \bm M_i/M$ is the unit vector aligned with the sublattice magnetization $\bm M_i$, $M$ is the sublattice saturation magnetization, $i=A,B$ is the sublattice index and $\gamma$ is the gyromagnetic ratio magnitude. $\alpha_{ij}$ is the $2 \times 2$ Gilbert dissipation matrix\cite{Kamra2018_2,Yuan2019}, which can be characterized by two real positive numbers $\alpha$ and $\alpha_c$ as follows: $\alpha_{AA} = \alpha_{BB}=\alpha$ and $\alpha_{AB}=\alpha_{BA}=\alpha_c$.  The last term in Eq.~\eqref{LLG} represents the torque experienced by the sublattice. 
$H_{eff}^i$ is the local effective field: \begin{eqnarray}
\bm H_{eff}^i = K m_{i,x}\bm e_x -K_\perp m_{i,y} \bm e_y + A\partial_x^2 \bm m_i - J\bm m_{\bar i},
\label{H_eff}
\end{eqnarray}
where the anisotropy easy (anisotropy constant $K$) and hard (anisotropy constant $K_\perp$) axes are taken along $x$ and $y$-directions, respectively. $A$ is the intra-sublattice exchange  stiffness and $J$ is the exchange coupling constant between the sublattices. 

The torque $\bm N_i$ can be calculated starting from the effective exchange interaction between the spin densities on the two sides of the S/AF interface~\cite{Kamra2017}:
\begin{eqnarray}
H_{int} = - \int d^2 \bm r J_{ex} \bm S_A \cdot \bm s,
\label{interface_ham}
\end{eqnarray}
where $\bm s$ is the electronic spin density operator in the S film, $\bm S_A$ is the localized spin operator in the AF film, belonging to the sublattice A. We assume that only the A sublattice is coupled to the interface, see Fig.~\ref{sketch}. $J_{ex}$ is the exchange constant and the integration is performed over the 2D interface.  It has been shown \cite{Kamra2018} that this exchange interaction hamiltonian results in the appearance of the exchange field $\bm h(\bm r) = -J_{ex} M \bm m_A (\bm r)/(2\gamma d_s)$ in the S film. 

Applying Ehrenfest's theorem from Eq.~(\ref{interface_ham}) one obtains the additional contribution to the Landau-Lifshitz-Gilbert equation written in the form of a torque acting on the magnetization:
\begin{eqnarray}
\bm N_A = J_{ex} \delta(z-z_I) \bm m_A (z) \times \langle \bm s \rangle,~~~ 
\bm N_B = 0,
\label{torque_general}
\end{eqnarray}
where the interface is located at $z=z_I$ and $\langle \bm s \rangle$ is the quantum mechanical averaged value of $\bm s$. Further we assume that the antiferromagnetic film is thin and its magnetization for a given sublattice is homogeneous in the $z$-direction. In this case Eqs.~(\ref{LLG}) and (\ref{torque_general}) can be averaged over the thickness $d_{AF}$ of the AF film. For the averaged torque we thus obtain
\begin{eqnarray}
\overline{\bm N}_A = \frac{J_{ex} \bm m_A \times  \langle \bm s\rangle}{d_{AF}},~~~ 
\overline{\bm N}_B = 0.
\label{torque_averaged}
\end{eqnarray}

\subsection{Microscopic calculation of the spin torque}
In order to find the magnetization  dynamics from Eq.~(\ref{LLG}) we need to calculate torque (\ref{torque_averaged}) microscopically by considering thermally-induced quantum transport mediated by Cooper pairs and quasiparticles in the superconductor. The detailed calculation of the spin torque for a given effective exchange field in the superconductor can be found in Ref.~\onlinecite{Bobkova2020} and is outlined below.

The superconductor is assumed to be in the ballistic limit. We neglect all the inelastic relaxation processes in the film assuming that its length is shorter than the corresponding relaxation length. As here we are dealing with the nonequilibrium problem, we work in the framework of the Keldysh technique for quasiclassical Green's functions. The matrix Green's function $\check g(\bm r, \bm p_F, \varepsilon, t)$ is a $8 \times 8$ matrix in the direct product of spin, particle-hole and Keldysh spaces and depends on the spatial vector $\bm r$, quasiparticle momentum direction $\bm p_F$, quasiparticle energy $\varepsilon$ and time $t$. In the S film it obeys the Eilenberger equation:
\begin{eqnarray}
 i \bm v_F \nabla\check g(\bm r, \bm p_F)+\Bigl[ \varepsilon \tau_z  + \bm h(\bm r) \bm \sigma \tau_z - \check \Delta,\check g \Bigr]_\otimes = 0,~~~~~~
 \label{eilenberger}
\end{eqnarray}
where $[A,B]_\otimes = A\otimes B -B \otimes A$ and $A \otimes B = \exp[(i/2)(\partial_{\varepsilon_1} \partial_{t_2} -\partial_{\varepsilon_2} \partial_{t_1} )]A(\varepsilon_1,t_1)B(\varepsilon_2,t_2)|_{\varepsilon_1=\varepsilon_2=\varepsilon;t_1=t_2=t}$. $\tau_{x,y,z}$ are Pauli matrices in particle-hole space with $\tau_\pm = (\tau_x \pm i \tau_y)/2$. $\hat \Delta = \Delta(x)\tau_+ - \Delta^*(x)\tau_-$ is the matrix structure of the superconducting order parameter $\Delta(x)$ in the particle-hole space.

In the ballistic limit treated here, it is convenient to use the so-called Riccati parametrization for the Green's function \cite{Eschrig2000,Eschrig2009}. In terms of the Riccati parametrization the retarded Green's function takes the form:
\begin{eqnarray}
\check g^{R,A} =
\pm N^{R,A} \otimes ~~~~~~~~~~~~~~~~~~~~~~~~~~~~ \nonumber \\ 
\left(
\begin{array}{cc}
1-\hat \Gamma^{R,A} \otimes \hat {\tilde \Gamma}^{R,A} & 2 \hat \Gamma^{R,A} \\
2 \hat {\tilde \Gamma}^{R,A} & -(1-\hat {\tilde \Gamma}^{R,A} \otimes \hat \Gamma^{R,A}) \\
\end{array}
\right),~~~~
\label{riccati_GF}
\end{eqnarray}

\begin{eqnarray}
\check g^{K} =
2N^R \otimes ~~~~~~~~~~~~~~~~~~~~~~~~~~~~~~~~~~~~~~~~~\nonumber \\
\left(
\begin{array}{cc}
x^K + \hat \Gamma^{R} \otimes \hat {\tilde x}^K \otimes \hat {\tilde \Gamma}^{A} & - (\hat \Gamma^{R} \otimes \hat {\tilde x}^K - \hat x^K \hat \Gamma^A) \\
\hat {\tilde \Gamma}^{R} \otimes \hat x^K - \hat {\tilde x}^K \otimes \hat {\tilde \Gamma}^{A}  & \hat {\tilde x}^K+\hat {\tilde \Gamma}^{R} \otimes \hat x^K \otimes \hat \Gamma^{A}) \\
\end{array}
\right) \otimes N^A ~~~~~~
\label{riccati_keldysh}
\end{eqnarray}
with
\begin{eqnarray}
N^{R,A} = \left(
\begin{array}{cc}
1+\hat \Gamma^{R,A} \otimes \hat {\tilde \Gamma}^{R,A} & 0 \\
0 & 1+\hat {\tilde \Gamma}^{R,A} \otimes \hat \Gamma^{R,A} \\
\end{array}
\right)^{-1}
\label{N}
\end{eqnarray}
where $\hat \Gamma^{R,A}$, $\hat {\tilde \Gamma}^{R,A}$, $\hat x^K$ and $\hat {\tilde x}^K$ are matrices in spin space. Note that our parametrization differs from the definition in the literature \cite{Eschrig2000,Eschrig2009} by factors $i\sigma_y$ as $\hat \Gamma_{standard}^{R,A} = \hat \Gamma^{R,A} i \sigma_y$ and $\hat {\tilde \Gamma}_{standard}^{R,A} = i \sigma_y \hat {\tilde \Gamma}^{R,A}$.
The Riccati parametrization Eq.~(\ref{riccati_GF}) obeys the normalization condition $\check g \otimes \check g = 1$ automatically.

The Riccati amplitude $\hat \Gamma$  obeys the following Riccati-type equations:
\begin{eqnarray}
 i \bm v_F \nabla \hat \Gamma^R + 2 \varepsilon \hat \Gamma^R = -\hat \Gamma^R \otimes \Delta^* \otimes \hat \Gamma^R - \bigl\{ \bm h \bm \sigma, \hat \Gamma^R \bigr\}_\otimes - \Delta ~~~~~~
 \label{riccati}
\end{eqnarray}
and $\hat {\tilde \Gamma}$ obeys the same equation with the substitution $\varepsilon \to -\varepsilon$, $\bm h \to -\bm h$ and $\Delta \to \Delta^*$.

The distribution function $\hat x^K$ obeys the equation:
\begin{eqnarray}
 i \bm v_F \nabla \hat x^K + i \partial_t \hat x^K + \hat \Gamma^R \otimes \Delta^* \otimes \hat x^K + \nonumber \\
 \hat x^K \otimes \Delta \otimes \hat {\tilde \Gamma}^A + [\bm h \bm \sigma,\hat x^K]_\otimes = 0 ,
 \label{distribution}
\end{eqnarray}
while $\hat {\tilde x}^K$ obeys the same equation with the substitution $\bm h \to -\bm h$, $\Delta \to \Delta^*$, $\hat \Gamma^{R,A} \leftrightarrow \hat {\tilde \Gamma}^{R,A}$. 

Considering a finite spatially inhomogeneous  magnetic texture like a domain wall, the Riccati amplitudes $\hat \Gamma$ and $\hat {\tilde \Gamma}$ can be found from Eq.~(\ref{riccati}) numerically with the following asymptotic condition:
\begin{align}
 \hat \Gamma_{\infty}  = & \Gamma_{0\infty} + \frac{\bm h_{\infty} \bm \sigma}{h} \Gamma_\infty , \nonumber \\
 \Gamma_{0\infty}  = & -\frac{1}{2}\Bigl[ \frac{\Delta}{\varepsilon +h+i\sqrt{\Delta^2 - (\varepsilon + h)^2}} \nonumber \\ 
 & + \frac{\Delta}{\varepsilon-h+i\sqrt{\Delta^2 - (\varepsilon  - h)^2}} \Bigr], \nonumber
 \\
 \Gamma_\infty = & -\frac{1}{2}\Bigl[ \frac{\Delta}{\varepsilon+h+i\sqrt{\Delta^2 - (\varepsilon  + h)^2}} \nonumber \\ & - \frac{\Delta}{\varepsilon -h+i\sqrt{\Delta^2 - (\varepsilon  - h)^2}} \Bigr],
  \label{riccati_asymptotic}
\end{align}
and $\hat {\tilde \Gamma}_\infty = - \hat \Gamma_\infty$. In Eqs.~(\ref{riccati_asymptotic}) $h = |\bm h|$ is the absolute value of the effective exchange field, which is spatially constant. $\varepsilon$ has an infinitesimal imaginary part \ak{$\delta$}, where $\delta$ is positive for the retarded functions.

Eq.~(\ref{riccati}) is numerically stable if it is solved starting from $x = -\infty$ for right-going trajectories $v_x > 0$ and from $x = +\infty$ for left-going trajectories $v_x < 0$. On the contrary, $\hat {\tilde \Gamma}$ can be found numerically starting from $x = +\infty$ for right-going trajectories $v_x > 0$ and from $x = -\infty$ for left-going trajectories $v_x < 0$. The advanced Riccati amplitudes can be found taking into account the relation~\cite{Eschrig2009} $\hat \Gamma^A = -(\hat {\tilde \Gamma}^R)^\dagger$. 

If we neglect the dependence of $\bm h$ on time, then it follows from Eq.~(\ref{distribution}) that the distribution function $\hat x^K$ for a given ballistic trajectory is determined by the equilibrium distribution function  of the left (right) reservoir for $v_{F,x} >0$ ($v_{F,x} <0$) and takes the form
\begin{eqnarray}
\hat x^K_{\pm} = (1+\hat \gamma^R_\pm \otimes \hat {\tilde \gamma}^A_\pm)\tanh \frac{\varepsilon}{2T_{l,r}},
\label{distrib_eq}
\end{eqnarray}
where the subscript $+(-)$ corresponds to the trajectories $v_{F,x}>0$ ($v_{F,x}<0$). On the contrary,
\begin{eqnarray}
\hat {\tilde x}^K_{\pm} = -(1+\hat {\tilde \gamma}^R_\pm \otimes \hat \gamma^A_\pm)\tanh \frac{\varepsilon}{2T_{r,l}}.
\label{tilde_distrib_eq}
\end{eqnarray}
The terms $\propto \dot {\bm h}$ in Eq.~(\ref{distribution}) can be neglected under the condition $(h/\Delta)v_{st}/l_{DW}\Delta \ll 1$, where $v_{st}$ is the velocity of the rigid DW motion caused by the thermal gradient under consideration, and $l_{DW}$ is the DW width. For realistic parameters $v_{st} \sim 100 m/s$ according to our estimates below. Therefore, at $\Delta \sim 1K$ and $h/\Delta \lesssim 1$ these conditions are fulfilled to a good accuracy for any experimentally reasonable DW width $l_{DW} \sim 10nm - 1\mu m$.

The superconducting order parameter is found self-consistently according to
\begin{eqnarray}
\Delta = -\frac{\lambda}{8} \int \limits_{-\Omega}^\Omega d \varepsilon {\rm Tr}_4\langle \tau_- \check g^K \rangle,
\label{self_con}
\end{eqnarray}
where $\langle ... \rangle$ denotes averaging over the Fermi surface, $\lambda$ is the coupling constant and $\Omega$ is the Debye frequency cutoff. The spatial dependence of the superconducting order parameter due to the localized domain wall is found to be weak~\cite{Bobkova2020}. On the other hand, the suppression of the order parameter due to finite temperature and exchange field in the superconductor is relatively important. Therefore, in the present study we only account for the spatially uniform temperature and exchange field-induced suppression of superconductivity neglecting the tiny spatial effects near the DW. In this case the order parameter can be calculated using the bulk expressions for the Riccati amplitudes Eqs.~(\ref{riccati_asymptotic}). Substituting the Riccati amplitudes and the distribution functions (\ref{distrib_eq}), (\ref{tilde_distrib_eq}) into the self-consistency equation (\ref{self_con}) we finally end up with
\begin{eqnarray}
\Delta = -\frac{\lambda}{4} \int \limits_{0}^\Omega d \varepsilon {\rm Re} \Bigl[ \frac{i\Delta}{\sqrt{\Delta^2 - (\varepsilon+i\delta+h)^2}}+ \nonumber \\
\frac{i\Delta}{\sqrt{\Delta^2 - (\varepsilon+i\delta-h)^2}} \Bigr]  
\Bigl( \tanh \frac{\varepsilon}{2T_l}+\tanh \frac{\varepsilon}{2T_r} \Bigr).
\label{self_con_bulk}
\end{eqnarray}

From Eq.~(\ref{eilenberger}) it can be shown that  $ \langle \bm s \rangle$ obeys the following equation:
\begin{eqnarray}
\partial_t \langle \bm s \rangle = - \partial_j \bm J_j - 2 \bm h \times \langle \bm s \rangle,
\label{electron_spin}
\end{eqnarray}
where we have introduced vector $\bm J_j = (J_j^x,J_j^y,J_j^z)$ corresponding to the spin current flowing along the $j$-axis in the coordinate space:
\begin{eqnarray}
\bm J_j = -\frac{ N_F}{16} \int \limits_{-\infty}^\infty d \varepsilon {\rm Tr}_4 \Bigl[ \bm \sigma \langle v_{F,j} \check g^K \rangle \Bigr],
\label{spin_current}
\end{eqnarray}
where $N_F$ is the normal state density of states at the Fermi level and $v_F$ is the Fermi velocity.  

Considering the steady state of the conduction electrons,  Eq.~(\ref{electron_spin}) yields
\begin{eqnarray}
\overline{\bm N}_A = \frac{\gamma d_S}{M d_{AF}} \partial_j \bm J_j.
\label{torque}
\end{eqnarray}

\section{Results}

\subsection{Numerical evaluation of thermally-induced DW motion}

Now we present the results of numerical simulations of the magnetization dynamics in the AF based on LLG Eqs.~(\ref{LLG}). The spin torque exterted by the superconductor is calculated microscopically according to Eqs.~(\ref{spin_current})-(\ref{torque}). The Neel vector  in the AF $\bm n = (\bm m_A - \bm m_B)/2$ can be parametrized as 
\begin{eqnarray}
\bm n = (\cos \theta, \sin \theta \sin \phi, \sin \theta \cos \phi),
\label{parametrization}
\end{eqnarray}
where both angles $\theta$ and $\phi$ depend on the $x$-coordinate. The equilibrium shape of the DW in the absence of the superconducting film is given by $
 \cos\theta = \tanh (x/d_{DW})$ 
and $\phi = 0$, that is the DW is in the x-z plane. 

For the numerical calculation we introduce the dimensionless quantities $\tilde t = t (\gamma K)$ and $\tilde {\bm H}_{eff} = m_{i,x}\bm e_x -k m_{i,y} \bm e_y + \tilde A\partial_x^2 \bm m_i - \tilde J\bm m_{\bar i}$ with $\tilde A = A/K$, $\tilde J = J/K$, $k=K_\perp/K$. All lengths are measured in units of $\xi_S = v_F/\Delta_0$, $\tilde x = x/\xi_S$. Here $\Delta_0$ is the superconducting order parameter of the S film in the absence of the antiferromagnet at $T=0$. The dimensionless torque is $\tilde {\bm N}_A = \overline{\bm N}_A/\gamma K  = \zeta \partial_{\tilde x} \tilde {\bm J}_x$, where the dimensionless quantity $\partial_{\tilde x} \tilde {\bm J}_x = (2e^2 R_N v_F/\Delta_0^2) \partial_x \bm J_x$ and $\zeta = E_S/\pi E_A$ is proportional to the ratio of the condensation energy $E_S = N_F \Delta_0^2 d_S/2$ and the anisotropy energy $E_A = M K d_{AF}/2$. Here and below $R_N = \pi/(2e^2 N_F v_F)$ is the normal state resistance of the S film.  For estimates we take $E_S \sim  (10-10^3)\times d_S$ $erg/sm^2$ (for conventional superconductors like Al and Nb)  and assume that characteristic values of $E_A$ for antiferromagnets are close to the corresponding values for ferromagnets: $E_A \sim 10^5 d$ $erg/sm^2$ for Py thin films \cite{Beach2006} or $E_A \sim (10-10^2)\times d$ $erg/sm^2$ for YIG thin films \cite{Mendil2019}. This implies that $\zeta$ can vary in wide range $\zeta \sim (10^{-4}-10^2)(d_S/d_{AF})$. For our numerical analysis, we assume $\zeta = 0.048$.

\begin{figure}[tb]
 \begin{minipage}[b]{\linewidth}
   \centerline{\includegraphics[clip=true,width=3.0in]{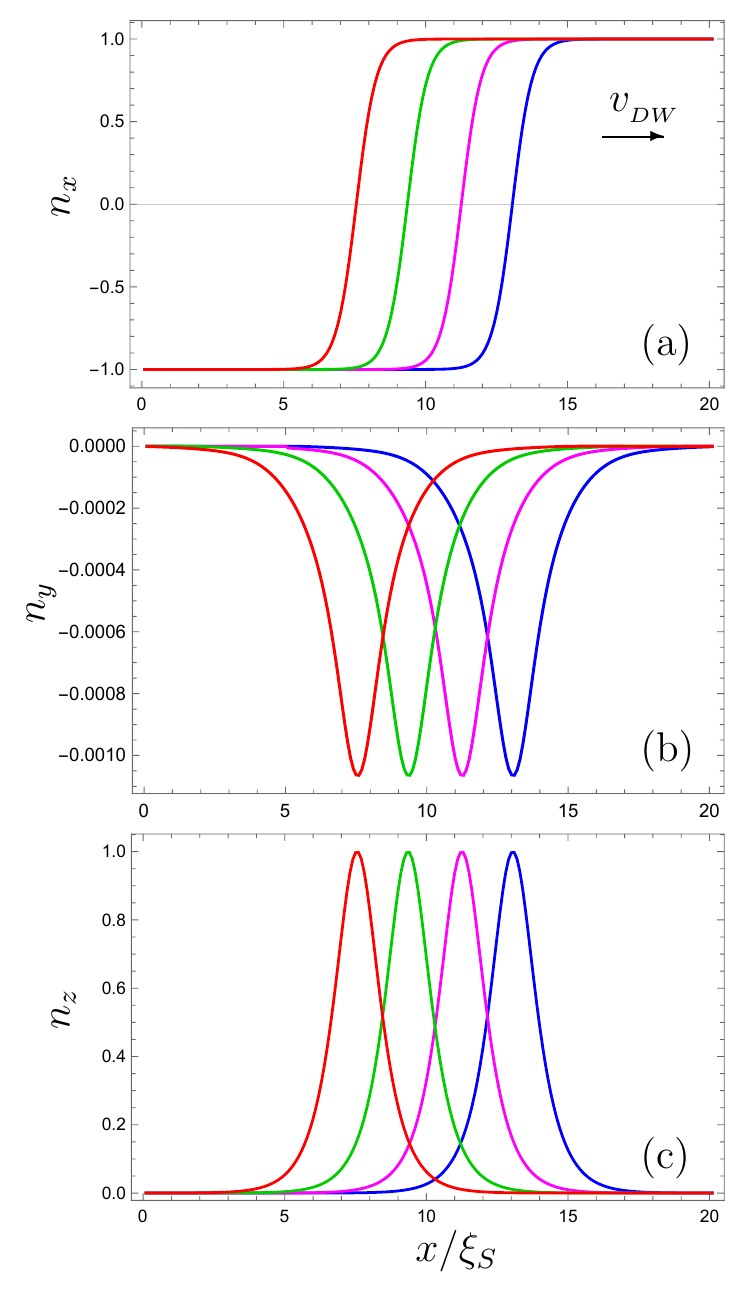}}
   \end{minipage}
      \caption{Spatial profiles of the Neel vector at several subsequent times. $T_l =0.32 \Delta_0$, $T_r =0.02 \Delta_0 $, $d=\xi_S$, $\alpha =0.01 $, $\alpha_c = 0.009$, $h=0.3\Delta_0$, the time between two subsequent curves $dt = 10t_0$. The DW moves from the left (hot) to the right (cold) end. The direction of the DW motion is indicated by the arrow in panel (a).}
 \label{profiles1}
 \end{figure}
 
\begin{figure}[!tbh]
 \begin{minipage}[b]{\linewidth}
   \centerline{\includegraphics[clip=true,width=3.0in]{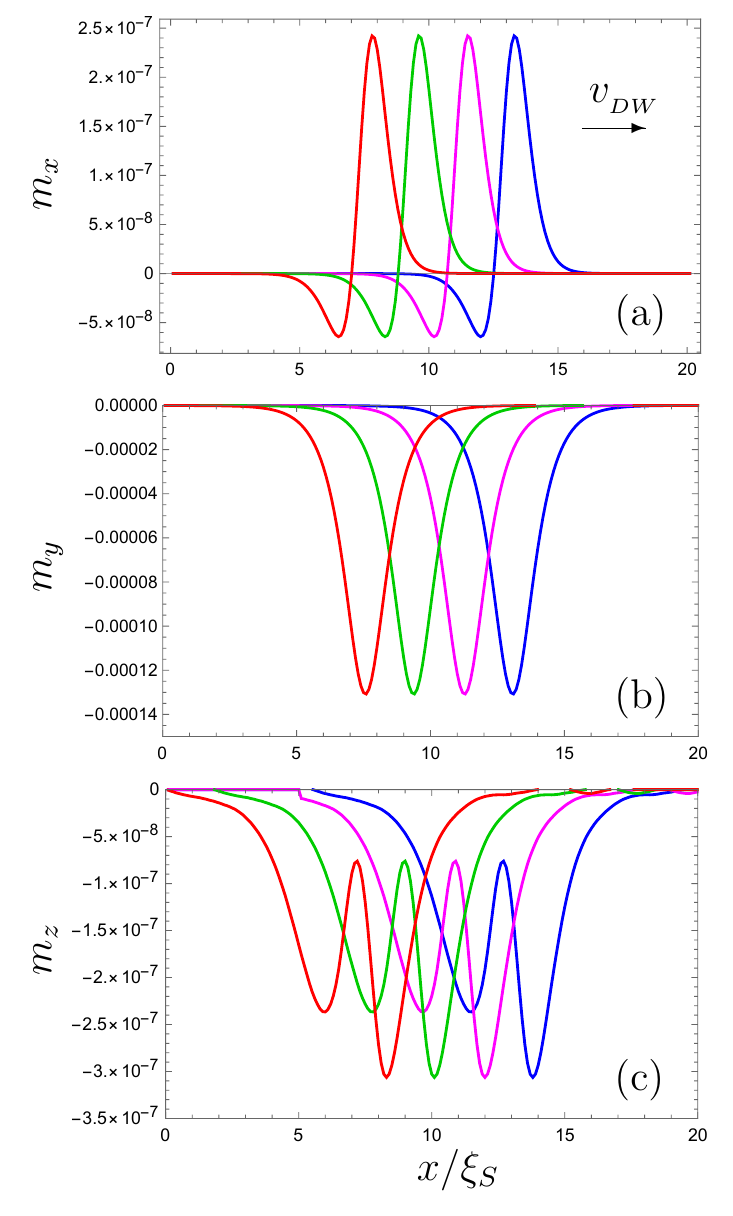}}
   \end{minipage}
      \caption{Spatial profiles of the magnetization at several subsequent times. The parameters are the same as in Fig.~\ref{profiles1}.}
 \label{profiles2}
 \end{figure}

Figs.~\ref{profiles1} and \ref{profiles2} depict snapshots of the spatial profiles of the Neel vector $\bm n $ and the magnetization $\bm m = (\bm m_A + \bm m_B)/2$ of the AF DW texture at several subsequent moments during the DW motion under the applied temperature difference for the finite value of the hard-axis anisotropy $k=1$. It is seen that the DW moves as a rigid object preserving its initial shape, that is the motion is in the regime well below the Walker breakdown~\cite{Schryer1974}. The motion is in this regime for the entire temperature range where the superconductivity survives. It is worth noting that the magnetization component $m_y$ is in $1/(\alpha-\alpha_c)$ times larger that the other two components, in agreement with the analytical treatment discussed below.

\begin{figure}[tbh]
 \begin{minipage}[b]{\linewidth}
   \centerline{\includegraphics[clip=true,width=3.0in]{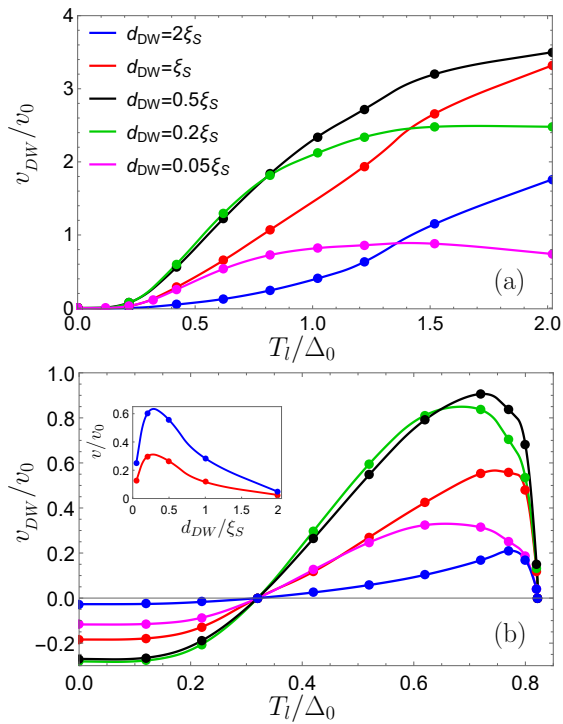}}
   \end{minipage}
      \caption{(a) DW velocity as a function of $T_l$ at $T_r = 0.02 \Delta_0$ for different DW widths.  (b) DW velocity as a function of $T_l$ at $T_r = 0.32\Delta_0$ for different DW widths. The inset shows the DW velocity at $T_l = 0.42\Delta_0$ and $T_r = 0.32\Delta_0$(red) or $T_r =0 $(blue) as a function of the DW width. The parameters $\alpha$, $\alpha_c$ and $h$ are the same as in Fig.~\ref{profiles1}.}
 \label{velocity}
 \end{figure}

In Fig.~\ref{velocity} we plot the velocity of the moving DW as a function of the left (hot) end temperature $T_l$. The velocity is measured in units of $v_0 = \xi_S/t_0$, where $t_0 = (\gamma K)^{-1}$. Taking typical values of the superconducting Al coherence length $\xi_S \sim 200$nm and $K \sim 10^2$G we can roughly estimate $v_0 \sim 10^4 - 10^5$cm/s. Fig.~\ref{velocity}(a) demonstrates the DW velocity for nearly zero temperature of the right end of the bilayer $T_r=0.02\Delta_0$. It is seen that the velocity becomes finite even at $T_l>T_{c0} \approx 0.57\Delta_0$, where $T_{c0}$ is the critical temperature of the superconductor in the absence of the antiferromagnet. This implies that the superconductivity still survives at such temperature differences. This existence of superconductivity at large temperatures is a specific feature of the ballistic limit, which results from the fact that at a given point only a half of all the trajectories, corresponding to $v_x>0$ carry hot quasiparticles distributed in accordance with $T_l$. The other half of trajectories $v_x<0$ carry no quasiparticles because they are not produced at the right end at $T=0$. Mathematically these arguments are expressed by Eq.~(\ref{self_con_bulk}), where the order parameter is determined by the sum of two Fermi functions, corresponding to the both ends of the bilayer.

At the same time Fig.~\ref{velocity}(b) corresponds to $T_r = 0.32\Delta_0$. At this value of the cold end temperature the amount of left-moving qusiparticles from the right (cold) end is enough to completely  suppress superconductivity already at $T_l=0.82\Delta_0$. As a result the DW velocity goes to zero at this temperature.

Different curves in Fig.~\ref{velocity} correspond to different values of the DW width $d_{DW}$ in units of $\xi_S$. It is seen that for a given temperature difference the DW velocity is a nonmonotonic function of $d_{DW}$ exhibiting a maximum at $d_{DW} \sim \xi_S/3$. The dependence of the DW velocity as a function of $d_{DW}$ is demonstrated in the inset of Fig.~\ref{velocity}(b). This can be understood via the argument that the DW motion is driven by the nonadiabatic torque component, which is $\propto d_{DW}/\xi_S$ at small values of this parameter (see analytical calculations below). Furthermore, the spin torque vanishes at large $d_{DW} \gg \xi_S$ because the electron spin can trace the magnetization in this case and the nonadiabatic torque goes to zero. Thus, the spin torque exerted is bound to achieve a maximum value in between the two extremes where it vanishes.

\subsection{Analytical calculation of the DW velocity in the framework of the collective coordinates approach}

For analytical calculations we substitute $ \bm m_{A,B} = \bm m \pm \bm n $ into Eq.~(\ref{LLG}) and after some algebra obtain:
\begin{eqnarray}
\frac{\partial \bm n}{\partial t} = -\gamma \Bigl[ \bm n \times \bm H^m + \bm m \times \bm H^n \Bigr] + \frac{\overline{\bm N}_A}{2}+ \nonumber \\
\alpha_n \bm n \times \frac{\partial \bm m}{\partial t} + \alpha_m \bm m \times \frac{\partial \bm n}{\partial t},
\label{LLG_n}
\\
\frac{\partial\bm m}{\partial t} = -\gamma \Bigl[ \bm n \times \bm H^n + \bm m \times \bm H^m \Bigr] + \frac{\overline{\bm N}_A}{2}+ \nonumber \\
\alpha_m \bm m \times \frac{\partial \bm m}{\partial t} + \alpha_n \bm n \times \frac{\partial \bm n}{\partial t}
\label{LLG_m}
\end{eqnarray}
where $\bm H^{m,n} = (\bm H_{eff}^A \pm \bm H_{eff}^B)/2$ and $\alpha_{m,n} = \alpha \pm \alpha_c$. We further take into account that $m \ll 1$. Then multiplying Eq.~(\ref{LLG_n}) by $\times \bm n$ and accounting for $n^2 \approx 1$ and $\bm n \cdot \bm m = 0$, we obtain the leading order expression for $\bm m$:
\begin{eqnarray}
\bm m = \frac{1}{2 \gamma J} \Bigl[ \partial_t \bm n \times \bm n + \bm n \times \frac{\overline{\bm N}_A}{2} \Bigr],
\label{m}
\end{eqnarray}
where we have also neglected small terms $\sim \alpha_{m,n}$ and $\alpha_{m,n}^2$. Eq.~(\ref{m}) is quite standard~\cite{Gomonay2014} except for the fact that the torque term, in our case, stems only from one of the sublattices. 

When magnetic textures are rigid, only a few soft modes dominate the magnetization dynamics. In this case the evolution of the soft modes can be described by a finite set of collective coordinates. This method was successfully applied both for ferromagnetic \cite{Schryer1974,Tretiakov2008,Clarke2008} and antiferromagnetic  \cite{Tveten2013} textures. 
Our numerical results presented in Fig.~\ref{profiles1} demonstrate that the magnetization texture for the problem under consideration is rigid and, therefore, we  exploit the collective coordinate method to analytically describe the DW motion. We use the DW center coordinate $x_{DW}$ and the out-of-plane tilt angle $\phi$ as collective coordinates. In this case the Neel vector of the moving DW can be written in the form of Eq.~(\ref{parametrization})
with $\theta = \theta[x-x_{DW}(t)]$ and $\phi=\phi(t)$. We consider the regime of a stationary motion of the DW and in this case $\phi(t) =\phi_0$. 

Keeping in Eq.~(\ref{LLG_m}) only terms up to the linear order with respect to $\overline{\bm N}_A$ we can write
\begin{eqnarray}
 -\gamma \bm n \times \bm H^n + \alpha_n \bm n \times \dot{\bm n} + \frac{\overline{\bm N}_A}{2} = 0.
\label{LLG_m_approx}
\end{eqnarray}
By projecting this equation on the $y$-axis and substituting the Neel vector in the form (\ref{parametrization}) we get
\begin{eqnarray}
 \alpha_n \theta' \dot x_{DW} + \frac{\overline{N}_{A,y}^{ne}}{2} = 0,
\label{LLG_m_approx_2}
\end{eqnarray}
where $\theta'$ is the derivative of $\theta$ with respect to its argument and $\overline{N}_{A,y}^{ne}=\overline{N}_{A,y}-\overline{N}_{A,y}^{eq}$ is the nonequilibrium part of the torque, which arises due to applied temperature difference. The equilibrium contribution $\overline{N}_{A,y}^{eq}$ exists also at $\Delta T =0$ and is compensated by a small distortion of the DW shape in S/AF bilayer with respect to the isolated AF film~\cite{Bobkova2020}. This equation is strictly valid only for the special shape of the torque $\overline{\bm N}_{A,y}^{ne} \propto \theta'$. Our numerical calculations indicate that this condition is approximately valid at $|x-x_{DW}| < x_0$, where $x_0 \lesssim \xi_S$. Therefore, the DW velocity can be approximately found as
\begin{eqnarray}
v_{DW} = \dot x_{DW} = \frac{\overline{N}_{A,y}^0 d_{DW}}{2 \alpha_n},
\label{v_local}
\end{eqnarray}
where $\overline{N}_{A,y}^0  = \overline{N}_{A,y}(x=x_{DW})$. A slightly more accurate result taking into account the averaging over the DW region can be obtained by integrating Eq.~(\ref{LLG_m_approx_2}):
\begin{eqnarray}
 v_{DW} = \frac{\int \limits_{-x_0}^{x_0} \overline{ N}_{A,y}dx}{2\alpha_n [\theta(x_0)-\theta(-x_0)]}.
\label{v}
\end{eqnarray}
This integrated expression is not very useful for analytical calculation of the DW velocity in case $d_{DW} >\xi_S$ because the parameter $x_0$ can only be extracted from numerical calculations. At the same time here we focus on the regime $d_{DW} \ll \xi_S$, which is relevant for Al-based AF/S bilayers due to the relatively large coherence length $\xi_S$ in Al. In this regime the main part of the integral in Eq.~(\ref{v}) comes from the region $|x-x_{DW}|<d_{DW}$ and, therefore, the exact value of $x_0 \gg d_{DW}$ is not important.

Analogously, by taking a projection of Eq.~(\ref{LLG_m_approx}) on the $x$-axis one can find the tilt angle
\begin{eqnarray}
 \phi_0 = \frac{\int \limits_{-x_0}^{x_0} \overline{ N}_{A,x}dx}{4 d_{DW} \gamma K_\perp}.
\label{delta}
\end{eqnarray}
Comparing this result to the tilt angle for the ferromagnetic case
\begin{eqnarray}
\phi_{F}=\frac{1}{\gamma  K_\perp d_{DW}}\Bigl[ \frac{1}{2} \int \limits_{-x_0}^{x_0} dx N_x - \frac{1}{\pi \alpha} \int \limits_{-x_0}^{x_0} dx N_z\Bigr].~~~~
\label{delta_F}
\end{eqnarray}
we see that due to the absence of the last term $\propto \alpha^{-1} \gg 1$ in Eq.~(\ref{delta}) the tilt angle in antiferromagnets is much smaller than in ferromagnets and can be considered as a hard mode, as it has been indicated in Ref.~\onlinecite{Tveten2013}. It is a manifestation of the qualitatively different behavior of magnetization dynamics in ferromagnets and antiferromagnets.

\begin{figure}[tbh]
 \begin{minipage}[b]{\linewidth}
   \centerline{\includegraphics[clip=true,width=3.0in]{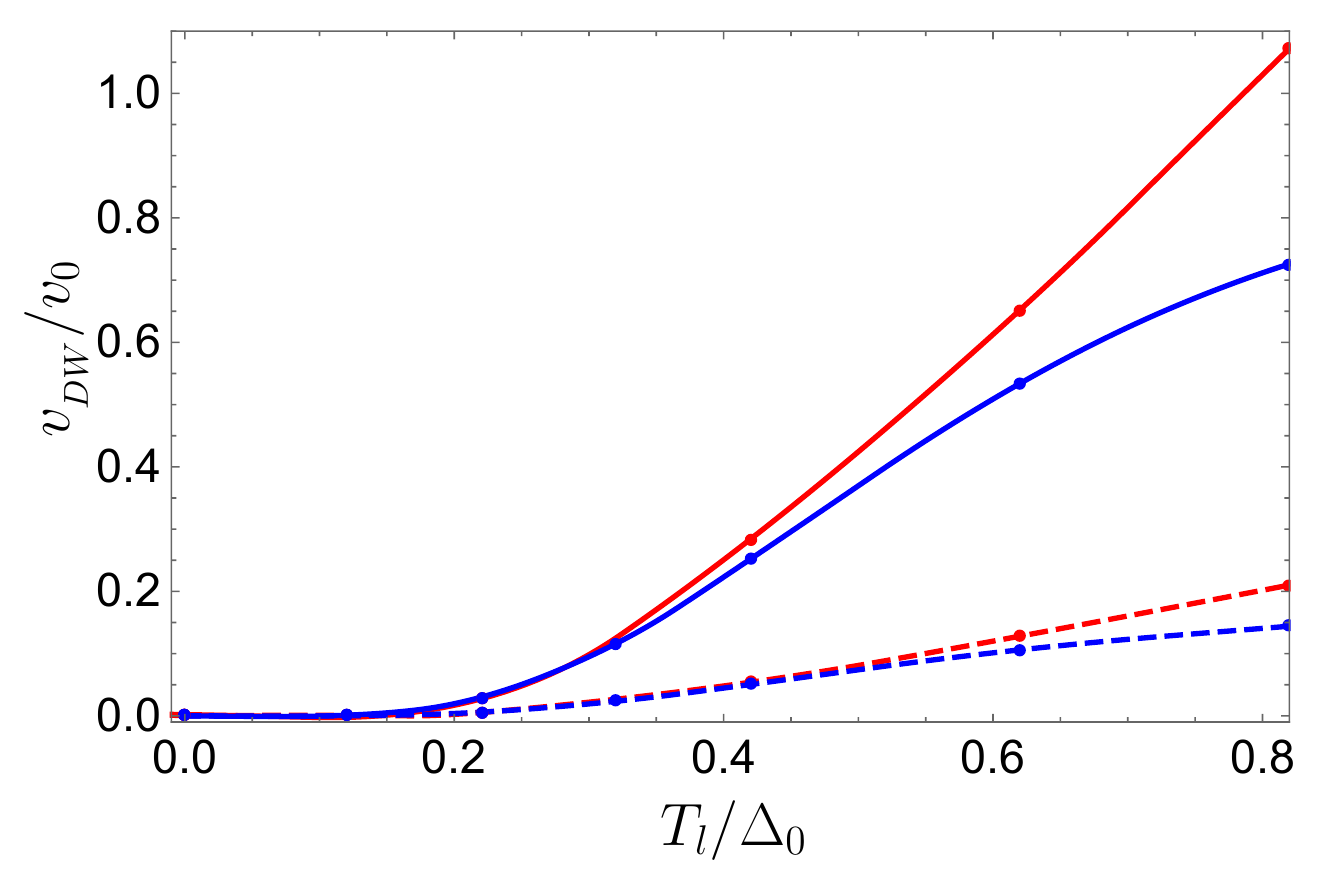}}
   \end{minipage}
      \caption{Numerically calculated DW velocity as a function of $T_l$ at different DW widths $d_{DW} = \xi_S$ (red) and $d_{DW} = 0.05 \xi_S$ (blue) and different values of $\alpha_n = 0.001$ (solid) and $\alpha_n = 0.005$ (dashed). $T_r = 0.02\Delta_0$. It is seen that the DW velocity is proportional to $\alpha_n^{-1}$ for any widths of the DW.}
 \label{velocity_alpha}
 \end{figure}

The DW velocity $v_{DW}$ is $\propto \alpha_n^{-1}$ in agreement with our numerical analysis (see Fig.~\ref{velocity_alpha}). It is also $\propto \int \limits_{-x_0}^{x_0} \overline{ N}_{A,y}dx$. For a plane DW under consideration the latter quantity is nothing but the nonadiabatic torque, integrated over the coordinate. Contrary to the phenomenological approaches, which were applied before to study  DW motion in ferromagnetic and antiferromagnetic textures, we calculate the nonadibatic torque microscopically. It is done numerically for a wide range of parameters and the resulting DW motion has been discussed in the previous section. 

In the regime $d_{DW} \ll \xi_S$ we are able to obtain an approximate analytical expression for the integrated nonadiabatic torque. According to Eq.~(\ref{torque}) $\int \limits_{-x_0}^{x_0} \overline{N}_{A,y} dx \propto \int \limits_{-x_0}^{x_0} d \bm J_y/dx = \bm J_y(x_0) - \bm J_y(-x_0)$. The last difference is mainly determined by the jump of the $y$-component of the spin current $\Delta \bm J_y$ at the DW. This jump can be found analytically in the framework of the perturbation theory with respect to the small parameter $d_{DW} /\xi_S \ll 1$. In this regime the DW can be viewed as a very narrow as compared to the superconducting coherence length and we can find the solution of Eq.~(\ref{riccati}) at the left and right boundaries of the DW, that is at $x=\mp d_{DW} \approx 0$, where the zero-order contribution is continuous, while the first order contribution to the Riccati-amplitudes exhibits a jump: 
\begin{eqnarray}
\hat \Gamma_+^R(-d_{DW}) = \hat \Gamma_{l,-\infty}^R, ~~~\hat \Gamma_{+}^{R}(d_{DW})=\hat \Gamma_{l,-\infty}^R+
\delta \Gamma \hat \sigma_z ,~~~
\label{gamma_1+}
\end{eqnarray}
\begin{eqnarray}
\hat \Gamma_-^R(d_{DW}) = \hat \Gamma_{r,+\infty}^R, ~~\hat \Gamma_{-}^{R}(-d_{DW})=\hat \Gamma_{r,+\infty}^R+
\delta \Gamma \hat \sigma_z ,~~~~
\label{gamma_1-}
\end{eqnarray}
\begin{eqnarray}
\hat {\tilde \Gamma}_+^R(d_{DW}) = -\hat \Gamma_{r,+\infty}^R, ~\hat {\tilde \Gamma}_{+}^{R}(-d_{DW})=-\hat \Gamma_{r,+\infty}^R-
\delta \Gamma \hat \sigma_z ,~~~~~
\label{tilde_gamma_1+}
\end{eqnarray}
\begin{eqnarray}
\hat {\tilde \Gamma}_-^R(-d_{DW}) = -\hat \Gamma_{l,-\infty}^R, ~\hat {\tilde \Gamma}_{-}^{R}(d_{DW})=-\hat \Gamma_{l,-\infty}^R-
\delta \Gamma \hat \sigma_z ,~~~~~
\label{tilde_gamma_1-}
\end{eqnarray}
where 
\begin{eqnarray}
\delta\Gamma=\int_{-\infty}^{+\infty}h_z (x) Tr[\hat \Gamma^{R,0}(x)]dx,
\label{delta_gamma}
\end{eqnarray}
where due to the condition $d_{DW}/\xi_S \ll 1$ the zero order contribution to the Riccati-amplitudes $\hat \Gamma^{R,0}(x)$ can be taken at $x=0$: $\hat \Gamma^{R,0}_{\pm}(x) \approx \Gamma^{R,0}_{\pm}(0) = \hat \Gamma_{l(r),\infty}^R$. Accounting for this approximation the first-order contribution to the Riccati-amplitudes takes the form:
\begin{eqnarray}
\delta \Gamma =\frac{2\pi d_{DW} h}{i |v_{F,x}|}\Gamma_{0\infty}
\label{delta_gamma_final}
\end{eqnarray}

Composing the Green's function from the Riccati amplitudes and substituting it into Eq.~(\ref{spin_current}) we end up with the following result:
\begin{eqnarray}
\Delta \bm J_y = \frac{\pi N_F d_{DW} h}{2} \int \limits_{-\infty}^\infty d \varepsilon  \frac{|I_1|^2-|I_2|^2}{|I_1 I_2 -1|^2} \times \nonumber \\
\Bigl[\tanh(\frac{\varepsilon}{2T_l})-\tanh(\frac{\varepsilon}{2T_r})\Bigr],
\label{jump}
\end{eqnarray} 
where 
\begin{eqnarray}
I_{1,2}=\frac{\varepsilon+i\delta \pm h + i\sqrt{\Delta^2-(\varepsilon+i\delta \pm h)2}}{\Delta}.
\end{eqnarray} 
Substituting this expression for $\Delta \bm J_y$ into Eq.~(\ref{v}) we finally obtain the following analytical expression for the DW velocity valid at $d_{DW} \ll \xi_S$:
\begin{eqnarray}
v_{DW} = Z^{-1}v_0 \frac{h}{\Delta_0}\int \limits_{-\infty}^\infty \frac{d \varepsilon}{\Delta_0}  \frac{|I_1|^2-|I_2|^2}{|I_1 I_2 -1|^2} \times \nonumber \\
\Bigl[\tanh(\frac{\varepsilon}{2T_l})-\tanh(\frac{\varepsilon}{2T_r})\Bigr],
\label{v_analytical}
\end{eqnarray} 
where $Z=8 \alpha_n \xi_S/(\pi \zeta d_{DW})$ is the parameter containing the dependence of the DW velocity on all the essential quantities, such as $\zeta$, $\alpha_n$ and $d_{DW}$, except for the dependence on the exchange field $h$.

\begin{figure}[!tbh]
 \begin{minipage}[b]{\linewidth}
   \centerline{\includegraphics[clip=true,width=3.0in]{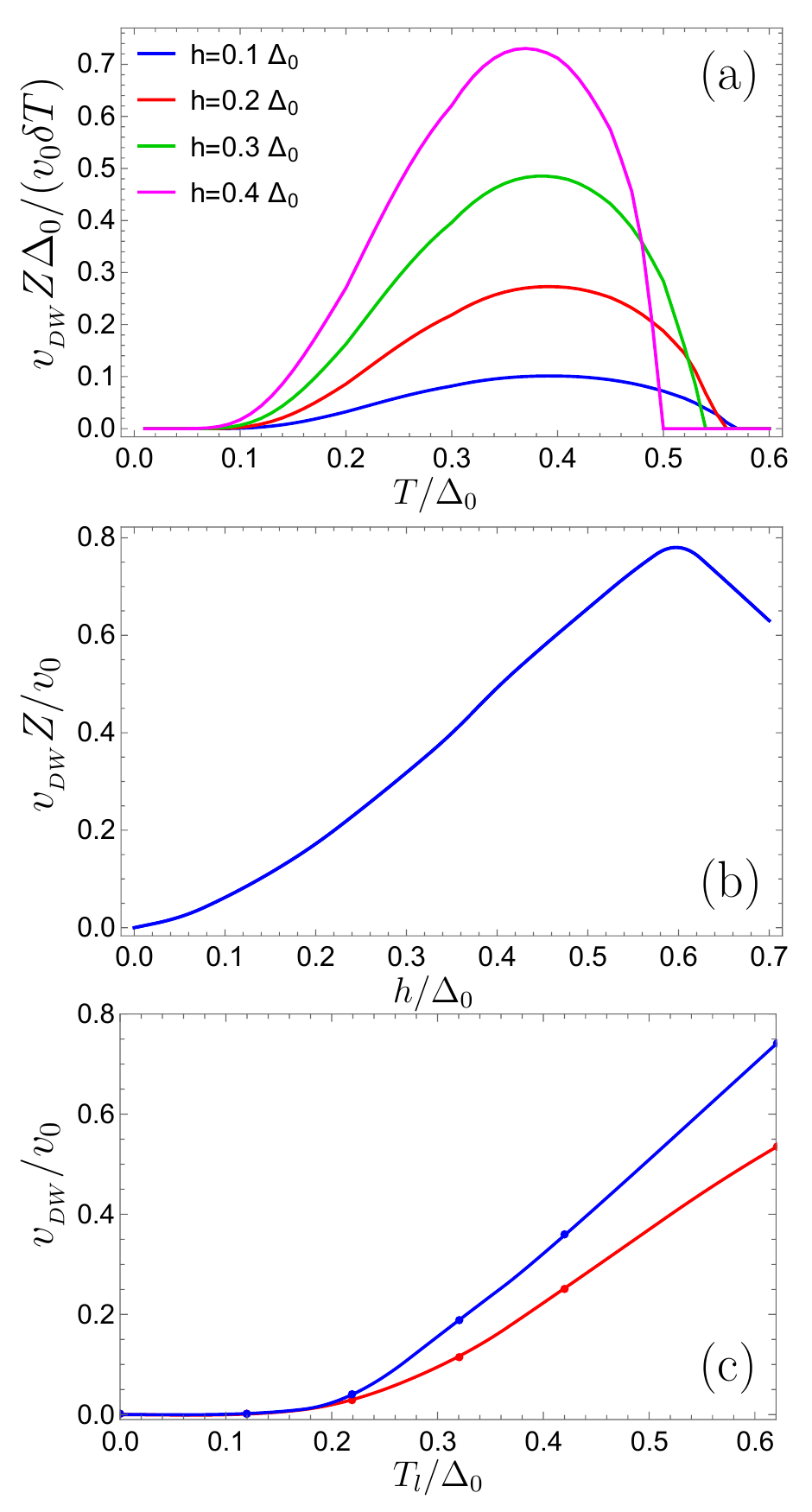}}
   \end{minipage}
      \caption{(a) $v_{DW}$ as a function of $T$ at small temperature differences $\delta T = T_l - T_r \ll T $ calculated according to the analytical expression Eq.~(\ref{v_analytical}). (b) Maximal $v_{DW}$ for a given $h$, which can be reached by properly adjusting the temperature difference $T_l - T_r$. (c) Comparison of the numerical (red) and analytical (blue) results for the DW velocity as a function of $T_l$. The cold end temperature $T_r = 0.02 \Delta_0$, $d_{DW} = 0.05 \xi_S$, $\alpha=0.01$, $\alpha_c=0.009$.}
 \label{velocity_analytical}
 \end{figure}

The DW velocity calculated according to Eq.~(\ref{v_analytical}) at small temperature differences $\delta T = T_l-T_r \ll T $ is shown in Fig.~\ref{velocity_analytical}(a) for different values of the exchange field $h$. It is seen that in general  $v_{DW}$ is higher for larger values of $h$, but it also more sharply vanishes at high temperatures because of the superconductivity suppression by the exchange field. Fig.~\ref{velocity_analytical}(b) demonstrates the maximum value of $v_{DW}$, which can be obtained for a given $h$ by properly adjusting the temperature difference $T_l-T_r$, calculated according to Eq.~(\ref{v_analytical}). The maximal value of the DW velocity grows with the exchange field until the superconductivity suppression by $h$ becomes strong enough and dominates in the dependence of $v_{DW}$ on the exchange field.

Eq.~(\ref{v_analytical}) can be further simplified at not very small values of the exchange field $0.1\Delta \lesssim h \lesssim \Delta$. In this case the integrand in Eq.~(\ref{jump}) can be approximated as
\begin{eqnarray}
\frac{|I_1|^2-|I_2|^2}{|I_1 I_2 -1|^2}\approx \frac{sign~\varepsilon}{4}(1+\frac{2\Delta}{h})\sqrt{\varepsilon^2-(\Delta-h)^2}
\label{intergand}
\end{eqnarray}
if $\varepsilon \in \pm [\Delta-h, \Delta+h]$ and it is zero beyond this energy interval. With this approximation 
\begin{eqnarray}
v_{DW} = Z^{-1}v_0 \frac{h}{\Delta_0}\Bigl[F(h,T_l) - F(h,T_r) \Bigr], \nonumber \\
F(h,T) = (1+\frac{2\Delta}{h})\sqrt{\frac{2(\Delta-h)}{\Delta_0}}\frac{T}{\Delta_0}e^{-\frac{\Delta-h}{T}}\times \nonumber \\
\Bigl( \frac{\sqrt{\pi}}{2} - \sqrt{\frac{2h}{T}}e^{-\frac{2h}{T}} \Bigr).
\label{v_analytical_2}
\end{eqnarray}
Eq.~(\ref{v_analytical_2}) can be further simplified at $T_{l,r} \ll \Delta$ resulting in:
\begin{eqnarray}
v_{DW} = \frac{\sqrt{\pi}}{2}Z^{-1}v_0 \frac{h}{\Delta_0}(1+\frac{2\Delta}{h})\sqrt{\frac{2(\Delta-h)}{\Delta_0}}\times \nonumber \\
\Bigl( T_l e^{-\frac{\Delta-h}{T_l}}-T_r e^{-\frac{\Delta-h}{T_r}} \Bigr).
\label{v_analytical_3}
\end{eqnarray}
Expressions (\ref{v_analytical_2}) and (\ref{v_analytical_3}) reflect the main qualitative features observed in the exact numerical results presented in Fig.~\ref{velocity}. In particular, $v_{DW}$ is exponentially suppressed at low temperatures $T_{l,r} \ll (\Delta-h)$, as it is seen in Fig.~\ref{velocity}(a) and can be qualitatively explained by the fact that the number of quasiparticles contributing to the giant thermospin effect is exponentially suppressed at such low temperatures. At moderate temperatures  $T > (\Delta-h)$ the DW velocity is roughly proportional to $T_l-T_r$, what is also seen from the numerical results. This behavior is changed by the velocity reduction upon further increase of temperature when the suppression of the superconducting gap by temperature becomes essential.

\section{Conclusions}
A high-efficiency thermally induced $180^\circ$ antiferromagnetic domain wall (DW) motion is predicted in thin-film AF/S hybrid structures with uncompensated magnetization at the AF/S interface. The surface magnetization gives rise to an effective exchange field and a spin splitting of the DOS in the superconductor. The physical reason of the torque providing the DW motion is  connected  to  the  generation  of  the  giant spin Seebek effect in the spin-split superconductor, which pumps quasiparticle spin into the superconducting region in  the  vicinity  of  the  DW.  The  pumped  spin  interacts with the AF interface magnetization via the interface exchange coupling. The resulting DW motion is investigated both numerically and analytically and the dependence of the DW velocity on the effective exchange field, Gilbert damping and the DW width is analyzed. Our estimates demonstrate that the suggested mechanism can lead to relatively high DW velocities $\sim 100m/s$ at small temperature differences $\sim 1K$ applied across a length equivalent to several domain wall widths.

\section*{Acknowledgements}

I.V.B. acknowledges the financial support by Foundation for the Advancement of Theoretical Physics and Mathematics “BASIS”.

\bibliography{STTRef.bib}

\end{document}